\documentclass[11pt]{article}

\usepackage{acl}

\usepackage{microtype}
\usepackage{times}
\usepackage{latexsym}

\usepackage{microtype}
\usepackage{float}

\usepackage[T1]{fontenc}
\usepackage[utf8]{inputenc}
\usepackage{microtype}
\usepackage{xurl}  % allows line breaks at any character in URLs

\usepackage{graphicx}
\usepackage{booktabs}
\usepackage{multirow}
\usepackage{amsmath}
\usepackage{amssymb}
\usepackage{xcolor}
\usepackage{tikz}
\usetikzlibrary{shapes.geometric, arrows.meta, positioning, fit, backgrounds, calc, decorations.pathreplacing}
\usepackage{pgfplots}
\pgfplotsset{compat=1.18}

\usepackage[ruled,vlined,linesnumbered]{algorithm2e}
\SetAlFnt{\small}
\SetAlCapFnt{\small}
\SetAlCapNameFnt{\small}

\usepackage{enumitem}
\setlist{nosep, leftmargin=*}

\usepackage{pifont}
\newcommand{\cmark}{\ding{51}}
\newcommand{\xmark}{\ding{55}}
\newcommand{\pmark}{$\sim$}

\title{AgentEval: DAG-Structured Step-Level Evaluation\\for Agentic Workflows with Error Propagation Tracking}

\author{
	Dongxin Guo$^{1}$, Jikun Wu$^{2}$, Siu Ming Yiu$^{1}$ \\
	$^{1}$The University of Hong Kong \quad $^{2}$Stellaris AI Limited \\
	\texttt{bettyguo@connect.hku.hk}, \texttt{hk950014@connect.hku.hk}, \\
	\texttt{smyiu@cs.hku.hk}
}

\hypersetup{
  pdftitle={AgentEval: DAG-Structured Step-Level Evaluation for Agentic Workflows with Error Propagation Tracking},
  pdfauthor={Dongxin Guo, Jikun Wu, Siu Ming Yiu},
  pdfsubject={ACL 2026 Industry Track},
  pdfkeywords={agent evaluation, LLM-as-judge, DAG, error propagation, root cause analysis, regression detection, agentic workflows}
}

\begin{document}
\maketitle

\begin{abstract}
	Agentic systems that chain reasoning, tool use, and synthesis into multi-step workflows are entering production, yet prevailing evaluation practices like end-to-end outcome checks and ad-hoc trace inspection systematically mask the intermediate failures that dominate real-world error budgets. We present \textsc{AgentEval}, a framework that formalizes agent executions as evaluation directed acyclic graphs (DAGs), where each node carries typed quality metrics assessed by a calibrated LLM judge (GPT-4o), classified through a hierarchical failure taxonomy (3 levels, 21 subcategories), and linked to upstream dependencies for automated root cause attribution. An ablation study isolates the impact of DAG-based dependency modeling: it alone contributes +22 percentage points to failure detection recall and +34~pp to root cause accuracy over flat step-level evaluation with identical judges and rubrics.
	
	Across three production workflows (450 test cases, two agent model families, predominantly sequential architectures with a 12\% non-DAG trace rate), \textsc{AgentEval} achieves $2.17\times$ higher failure detection recall than end-to-end evaluation ($0.89$ vs.\ $0.41$), Cohen's $\kappa = 0.84$ agreement with human experts, and 72\% root cause accuracy against an 81\% human ceiling. Cross-system evaluation on $\tau$-bench and SWE-bench traces confirms transferability (failure detection recall $\geq 0.78$) without taxonomy or rubric modification. A 4-month pilot with 18 engineers detected 23 pre-release regressions through CI/CD-integrated regression testing, reducing median root-cause identification time from 4.2 hours to 22 minutes and driving measurable failure rate reductions in two workflows. 
\end{abstract}

\section{Introduction}
\label{sec:intro}

Agentic AI systems (autonomous agents that plan, reason, and execute multi-step workflows using external tools) are transforming NLP applications. As of early 2025, 62\% of organizations are experimenting with AI agents and 23\% are scaling deployments \citep{mckinsey2025agents}, enabled by advances in agent frameworks \citep{wu2023autogen, schick2023toolformer, patil2024gorilla}. Yet evaluation infrastructure has not kept pace: current approaches rely on end-to-end outcome metrics that mask intermediate failures, ad hoc manual inspection that does not scale, or static benchmarks \citep{liu2024agentbench, jimenez2024swebench, zhou2024webarena} disconnected from deployment constraints such as latency, cost, and continuous integration. This evaluation gap has tangible consequences: Gartner predicts that over 40\% of agentic AI projects will be canceled by 2027, partly due to the inability to systematically evaluate deployed agents \citep{gartner2025cancellation}.

Multi-step agent workflows create evaluation challenges absent from traditional NLP evaluation. Agent executions follow DAG-structured dependencies where errors propagate and compound through downstream steps \citep{cemri2025multiagent, zhu2025agents}, and workflows involve heterogeneous step types each requiring distinct quality metrics. Process supervision research demonstrates that intermediate-step assessment outperforms outcome-only evaluation in the RL training setting where per-step ground truth is available \citep{lightman2024lets, uesato2022process}; we adapt this insight to inference-time assessment, where ground truth is constructed from expert annotation rather than learned from rewards, and where deployment infrastructure must operate continuously rather than as a one-time training signal.

We present \textsc{AgentEval},\footnote{The name ``AgentEval'' may have been used by other unrelated projects; this work is independent and identified by its GitHub repository (URL at the end of Section~\ref{sec:conclusion}).} an evaluation infrastructure addressing these challenges through four contributions:

\begin{enumerate}[leftmargin=1.5em]
\item \textbf{Evaluation DAG formalization}: representing agent workflows as DAGs where each node carries typed quality metrics, enabling step-level evaluation with error propagation tracking.
\item \textbf{Step-level quality metrics}: a metric suite evaluated via calibrated LLM-as-judge scoring (GPT-4o) with demonstrated human alignment ($\kappa = 0.84$).
\item \textbf{Hierarchical failure taxonomy}: a three-level taxonomy with 21 subcategories derived from 523 agent traces, with quantitative error propagation statistics.
\item \textbf{Automated regression suite}: evaluation infrastructure integrated with CI/CD pipelines for continuous quality monitoring.
\end{enumerate}

On three production workflows with predominantly sequential architectures (using Claude~3.5~Sonnet and Llama~3~70B as agents with GPT-4o as an independent judge), \textsc{AgentEval} achieves $2.17\times$ higher failure detection recall than end-to-end evaluation, $\kappa = 0.84$ human agreement, and 72\% root cause accuracy. We explicitly characterize the approach's boundaries for more dynamic architectures in \S\ref{sec:scope_analysis}.

\section{Related Work}
\label{sec:related}

\paragraph{Agent Evaluation and Process Supervision.}
Agent evaluation benchmarks have proliferated \citep{liu2024agentbench, jimenez2024swebench, zhou2024webarena, yao2024taubench, ma2024agentboard}, with LATS \citep{zhou2024lats} providing per-step value estimates for planning. However, a recent survey \citep{yehudai2025survey} identifies a critical gap: existing benchmarks evaluate agent \emph{capabilities} in controlled settings, whereas deployment requires evaluation \emph{infrastructure} with continuous monitoring, regression detection, and CI/CD integration. The process supervision literature demonstrates that evaluating intermediate steps outperforms outcome-only evaluation \citep{lightman2024lets, uesato2022process}, while \citet{cemri2025multiagent} and \citet{zhu2025agents} show that error propagation is the primary bottleneck in agent performance. The LLM-as-judge paradigm \citep{liu2023geval, zheng2023judging} enables scalable evaluation with $>$80\% human agreement.

\paragraph{Production Infrastructure.}
The ML observability ecosystem, including MLflow, Weights \& Biases Weave \citep{wandb2024weave}, LangSmith~\citep{langsmith2024}, Arize Phoenix~\citep{arize2024}, Braintrust~\citep{braintrust2024}, Inspect~\citep{inspect2024}, and AgentOps~\citep{agentops2024}, provides monitoring for ML pipelines and emerging agent support. These tools differ from \textsc{AgentEval} in one important way: none provides formal DAG-based dependency modeling with error propagation tracking and root cause attribution (Table~\ref{tab:tool_comparison}). \textsc{AgentEval}'s root cause attribution draws motivation from software engineering fault localization \citep{jones2005empirical, abreu2007accuracy}, operating as a practical greedy heuristic rather than formal causal inference.

\section{The \textsc{AgentEval} Framework}
\label{sec:framework}

\begin{figure*}[t]
	\centering
	\begin{tikzpicture}[
		scale=0.88, transform shape,
		node distance=0.45cm and 0.65cm,
		box/.style={rectangle, draw, rounded corners=2pt, minimum height=0.65cm, minimum width=1.6cm, font=\scriptsize, align=center, fill=#1},
		box/.default=blue!8,
		arrow/.style={-{Stealth[length=2.2pt]}, semithick},
		grouplabel/.style={font=\scriptsize\bfseries}
		]
		% Main pipeline
		\node[box=orange!12] (agent) {Agent\\System};
		\node[box=orange!12, right=of agent] (trace) {Trace\\Collector};
		\node[box=blue!12, right=of trace] (parser) {DAG\\Parser};
		\node[box=blue!12, right=of parser] (eval) {Step\\Evaluators};
		\node[box=blue!12, right=of eval] (agg) {Metric\\Aggregator};
		\node[box=green!12, right=of agg] (reg) {Regression\\Detector};
		\node[box=green!12, right=of reg] (cicd) {CI/CD\\Pipeline};
		
		% Auxiliary nodes
		\node[box=purple!10, below=0.5cm of parser] (rubric) {Evaluation\\Rubrics};
		\node[box=purple!10, below=0.5cm of eval] (tax) {Failure\\Taxonomy};
		\node[box=purple!10, below=0.5cm of agg] (dash) {Quality\\Dashboard};
		
		% Arrows
		\draw[arrow] (agent) -- (trace);
		\draw[arrow] (trace) -- (parser);
		\draw[arrow] (parser) -- (eval);
		\draw[arrow] (eval) -- (agg);
		\draw[arrow] (agg) -- (reg);
		\draw[arrow] (reg) -- (cicd);
		\draw[arrow] (rubric) -- (eval);
		\draw[arrow] (tax) -- (eval);
		\draw[arrow] (agg) -- (dash);
		
		% Background groups
		\begin{pgfonlayer}{background}
			\node[fit=(agent)(trace), draw=orange!50, fill=orange!3, rounded corners=4pt, inner xsep=5pt, inner ysep=10pt] (grp1) {};
			\node[grouplabel, text=orange!70!black, above=0pt of grp1.north] {Instrumentation};
			
			\node[fit=(parser)(eval)(agg)(rubric)(tax)(dash), draw=blue!50, fill=blue!3, rounded corners=4pt, inner xsep=5pt, inner ysep=10pt] (grp2) {};
			\node[grouplabel, text=blue!70!black, above=0pt of grp2.north] {Evaluation Engine};
			
			\node[fit=(reg)(cicd), draw=green!50, fill=green!3, rounded corners=4pt, inner xsep=5pt, inner ysep=10pt] (grp3) {};
			\node[grouplabel, text=green!70!black, above=0pt of grp3.north] {Integration};
		\end{pgfonlayer}
	\end{tikzpicture}
	\caption{\textsc{AgentEval} architecture. Agent traces are collected via OpenTelemetry-compatible instrumentation, parsed into evaluation DAGs, assessed by type-specific step evaluators using calibrated LLM-as-judge rubrics and the failure taxonomy, aggregated into workflow-level metrics, and fed into regression detection and CI/CD integration.}
	\label{fig:architecture}
\end{figure*}
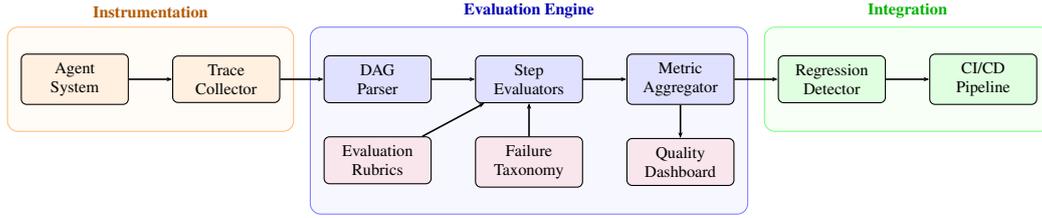

\textsc{AgentEval} comprises four integrated components: (1) a formal DAG representation for agent workflows, (2) step-level quality metrics evaluated via calibrated LLM-as-judge, (3) a hierarchical failure taxonomy, and (4) an automated regression suite. Figure~\ref{fig:architecture} provides an architectural overview.

\subsection{Agent Workflow as Evaluation DAG}
\label{sec:dag}

We formalize agent workflows as evaluation DAGs to enable structured step-level assessment.

\smallskip
\noindent\textbf{Definition 1} (Evaluation DAG). An evaluation DAG is a tuple $\mathcal{G} = (V, E, \tau, \mathcal{M})$ where $V = \{v_1, \ldots, v_n\}$ is a set of evaluation nodes corresponding to agent steps; $E \subseteq V \times V$ defines directed dependency edges; $\tau: V \rightarrow \mathcal{T}$ maps each node to a step type from $\mathcal{T} = \{\textsc{Plan}, \textsc{ToolSel}, \textsc{ParamGen}, \textsc{Exec}, \textsc{Synth}\}$; and $\mathcal{M}: V \rightarrow 2^{\mathbb{M}}$ maps each node to applicable quality metrics based on its type.

Each node $v_i$ carries input context $c_i$ from parent nodes $\text{pa}(v_i)$, agent output $o_i$, and optional reference $r_i$. The step-level quality score is $q(v_i) = \text{Eval}(o_i, r_i, c_i, \mathcal{M}(v_i))$. The five-type classification is extensible: the framework supports adding types (e.g., \textsc{MemRetrieval} for RAG agents) by defining new metric sets and rubrics.

The DAG structure enables independent step evaluation, error propagation tracking, counterfactual analysis, and parallel evaluation of independent branches. \textsc{AgentEval} supports both \emph{schema-defined DAGs} (expected workflow structure) and \emph{trace-inferred DAGs} (actual execution path), enabling detection of structural deviations such as missing steps or unexpected branches. Non-DAG traces ($\sim$12\%) are handled via loop unrolling and timestamp-based branch resolution, with 0.8\% falling back to flat evaluation (Appendix~\ref{sec:appendix_nondag}). During our pilot, structural deviation between schema-defined and trace-inferred DAGs proved a useful quality signal: deviant traces were associated with 2.1$\times$ higher failure rates than conformant traces. This is an observational pattern, not a causal claim; an agent may legitimately deviate from its expected DAG when the environment returns unexpected output, and deviation alone does not imply malfunction.

\subsection{Step-Level Quality Metrics}
\label{sec:metrics}

Each step type has dedicated metrics: \textsc{Plan} steps are assessed for completeness and feasibility; \textsc{ToolSel} for selection accuracy and relevance; \textsc{ParamGen} for parameter correctness and completeness; \textsc{Exec} for success rate and result validity; and \textsc{Synth} for faithfulness, completeness, and coherence. We employ LLM-as-judge evaluation using \textbf{GPT-4o} (\texttt{gpt-4o-2024-08-06}) with temperature $= 0$. The agent LLMs (Claude 3.5 Sonnet and Llama 3 70B) are from \emph{different model families} than the judge, addressing circular bias. Each metric uses a structured prompt with a 1--5 rubric, chain-of-thought reasoning, and 5 few-shot calibration anchors per metric (Appendix~\ref{sec:appendix_prompts}). By ``calibration'' we mean stratified few-shot anchoring with 5 examples spanning the score range, not iterative prompt optimization against a held-out human-labeled set.

\paragraph{Prompt design: absolute versus relative framing.} The judge prompts use two intentionally different framings depending on step type. \textsc{Plan} is scored relative to the original user query, since planning is the step that translates the query into workflow structure and has no meaningful upstream context (Appendix~\ref{sec:appendix_prompts}, C.2). Subsequent steps such as \textsc{ToolSel} are scored relative to the local context they received from upstream nodes rather than re-evaluated against the original query (Appendix~\ref{sec:appendix_prompts}, C.1). As a consequence of this asymmetry, a downstream step that handles flawed upstream input gracefully (e.g., a tool call that retries on a malformed parameter) will not be flagged as a local failure, while a step that compounds the error will. Downstream prompts include prior step outputs but not prior judge scores (Algorithm~\ref{alg:evaluation}, line 4), so judges do not bias toward agreement with upstream attributions.

\subsection{Hierarchical Failure Taxonomy}
\label{sec:taxonomy}

We derive a three-level failure taxonomy from manual analysis of 523 agent traces \textbf{entirely disjoint} from the 450 evaluation test cases (Table~\ref{tab:taxonomy}). The taxonomy is informed by prior work on multi-agent failures \citep{cemri2025multiagent, zhu2025agents}. The 9 Level~2 categories are used for detection and primary classification; 21 Level~3 subcategories (Appendix~\ref{sec:appendix_taxonomy}) provide granular diagnostic detail. Taxonomy classification runs as a separate LLM-judge call after a step has already been flagged as a failure by score thresholding; it produces a diagnostic label and does not interact with the propagation rule in Algorithm~\ref{alg:evaluation}. Removing the taxonomy (Table~\ref{tab:ablation}) therefore degrades RCA primarily through loss of semantic grouping during error analysis, not through changes to which steps are attributed as root cause.

\begin{table}[t]
\centering
\footnotesize
\begin{tabular}{@{}llr@{}}
\toprule
\textbf{Level 1} & \textbf{Level 2} & \textbf{Freq.} \\
\midrule
\multirow{3}{*}{Planning} & Goal misinterpretation & 12\% \\
 & Missing steps & 8\% \\
 & Incorrect ordering & 5\% \\
\midrule
\multirow{3}{*}{Execution} & Wrong tool selection & 18\% \\
 & Parameter errors & 22\% \\
 & API/tool failures & 9\% \\
\midrule
\multirow{3}{*}{Integration} & Context loss & 11\% \\
 & Output hallucination & 10\% \\
 & Premature termination & 5\% \\
\bottomrule
\end{tabular}
\caption{Hierarchical failure taxonomy derived from 523 agent traces (disjoint from evaluation data). Context loss has the highest downstream amplification factor (3.2$\times$).}
\label{tab:taxonomy}
\end{table}

\subsection{Automated Regression Suite}
\label{sec:regression}

The regression suite enables continuous quality monitoring through: test case specification as JSON with expected DAG structures and tolerance thresholds, versioned evaluation tagged with model/prompt/tool configurations, regression detection via paired bootstrap tests ($p < 0.05$) with dual-threshold alerting (historical $2\sigma$ \emph{and} bootstrap significance), CI/CD integration via GitHub Actions that blocks deployment on critical regressions, and progressive evaluation where fast smoke tests (10 cases, ${<}$5 minutes) gate full suites (100+ cases, ${<}$1 hour), reducing cost by 80\% during development.

Algorithm~\ref{alg:evaluation} details the core evaluation procedure. For each trace, the DAG is reconstructed, nodes are processed in topological order, and each step is evaluated with type-specific metrics via the LLM judge. Failures are classified using the taxonomy and attributed as either root-cause or propagated using a greedy heuristic: when multiple parents have low scores, the parent with the lowest quality score is selected as the propagation source. This prioritizes simplicity and low false attribution rates over formal causal reasoning (Appendix~\ref{sec:appendix_rca_strategies}).

\begin{algorithm}[t]
\SetAlgoLined
\KwIn{Agent execution trace $T$, DAG schema $S$ (optional)}
\KwOut{Evaluation report $R$ with step scores, failures, root causes}
$\mathcal{G} \leftarrow \text{ParseDAG}(T, S)$ \tcp*{Reconstruct DAG}
$\langle v_1, \ldots, v_n \rangle \leftarrow \text{TopologicalSort}(\mathcal{G})$\;
\For{$v_i$ \textbf{in} $\langle v_1, \ldots, v_n \rangle$}{
  $c_i \leftarrow \text{AggregateContext}(\text{pa}(v_i))$\;
  $q(v_i) \leftarrow \text{LLMJudge}(o_i, r_i, c_i, \mathcal{M}(\tau(v_i)))$\;
  \If{$q(v_i) < \theta_{\tau(v_i)}$}{
    $f_i \leftarrow \text{ClassifyFailure}(v_i, \text{Taxonomy})$\;
    \tcp{Greedy heuristic: select lowest-scoring parent}
    \eIf{$\exists\, v_j \in \text{pa}(v_i): q(v_j) < \theta_{\tau(v_j)}$}{
      $v_j^* \leftarrow \arg\min_{v_j \in \text{pa}(v_i)} q(v_j)$\;
      Mark $v_i$ as \emph{propagated} from $v_j^*$\;
    }{
      Mark $v_i$ as \emph{root cause}\;
    }
  }
}
$R \leftarrow \text{Aggregate}(\mathcal{G}, \{q(v_i)\}, \{f_i\})$\;
\Return{$R$}\;
\caption{\textsc{AgentEval} Evaluation Procedure}
\label{alg:evaluation}
\end{algorithm}

\section{Experimental Setup}
\label{sec:setup}

\subsection{Agent Workflows}

We evaluate on three agent workflows deployed in customer service (CS), data analysis (DA), and document processing (DP) pipelines (Table~\ref{tab:workflow_specs}). Each uses the same tool set and prompt templates as its production counterpart but operates on anonymized historical queries.

\begin{table}[t]
\centering
\footnotesize
\setlength{\tabcolsep}{2.5pt}
\begin{tabular}{@{}lccc@{}}
\toprule
\textbf{Specification} & \textbf{CS} & \textbf{DA} & \textbf{DP} \\
\midrule
Agent LLM & \multicolumn{2}{c}{Claude 3.5 Sonnet} & Llama 3 70B \\
Framework & \multicolumn{2}{c}{LangChain} & Custom \\
\# Tools available & 8 & 6 & 5 \\
Avg.\ steps/trace & 6.1 & 5.2 & 3.8 \\
\# Test cases & 150 & 150 & 150 \\
Edge case \% & 22\% & 18\% & 15\% \\
\bottomrule
\end{tabular}
\caption{Workflow specifications. Agent LLMs are from different model families than the GPT-4o judge.}
\label{tab:workflow_specs}
\end{table}

\subsection{Evaluation Dataset}
\label{sec:dataset}

We use two \textbf{completely disjoint} datasets: a \emph{taxonomy development set} of 523 traces (January--March 2025), and an \emph{evaluation dataset} of 150 test cases per workflow (450 total, April--June 2025), sampled using stratified random sampling ($\sim$80\% typical, $\sim$20\% edge cases). Ground truth annotations were produced by 5 domain experts (3+ years experience), \emph{none of whom are paper authors}. Each test case is a multi-step trace, and annotation is performed at the step level: the human-evaluation subset of 150 traces yields 987 step annotations in total, each rated 1--5 on the type-specific rubric. Of these, 195 steps are rated $\leq$2 by the human majority and constitute the failing-step set used as the Failure Detection Recall denominator (Section~\ref{sec:dataset}; full step distribution in Appendix~\ref{sec:appendix_confusion}). For human evaluation, 9 separate domain experts independently rated 150 cases on a 1--5 rubric (inter-annotator $\kappa = 0.79$, Fleiss' $\kappa = 0.76$). Root causes were independently annotated with inter-annotator $\kappa_{\text{RCA}} = 0.74$ (81\% pairwise agreement), which we treat as the human ceiling.

\subsection{Baselines and Metrics}

\textbf{End-to-End Only (E2E)}: evaluates only final output quality using GPT-4o with no intermediate assessment. \textbf{Flat Step Evaluation (Flat)}: evaluates each step independently using identical GPT-4o judge and rubrics but without DAG structure. \textbf{Rule-Based}: 47 hand-crafted deterministic rules across the three workflows.

We measure \textbf{Failure Detection Recall (FDRec)}, \textbf{False Positive Rate (FPR)}, \textbf{Human Agreement (HA, Cohen's $\kappa$)}, and \textbf{Root Cause Accuracy (RCA)}. All metrics reported with 95\% bootstrap CIs (10,000 resamples).

\section{Results and Analysis}
\label{sec:results}

\subsection{Main Results}

\begin{table}[t]
\centering
\footnotesize
\setlength{\tabcolsep}{3pt}
\begin{tabular}{@{}lcccc@{}}
\toprule
\textbf{Method} & \textbf{FDRec}$\uparrow$ & \textbf{FPR}$\downarrow$ & \textbf{HA} ($\kappa$)$\uparrow$ & \textbf{RCA}$\uparrow$ \\
\midrule
E2E Only & .41\tiny{ [.35,.47]} & .08\tiny{ [.05,.12]} & .52\tiny{ [.45,.58]} & N/A \\
Flat Step & .67\tiny{ [.61,.73]} & .15\tiny{ [.11,.20]} & .71\tiny{ [.65,.76]} & .38\tiny{ [.30,.46]} \\
Rule-Based & .58\tiny{ [.52,.64]} & \textbf{.05}\tiny{ [.03,.08]} & .63\tiny{ [.57,.69]} & .45\tiny{ [.37,.53]} \\
\textsc{AgentEval} & \textbf{.89}\tiny{ [.84,.93]} & .07\tiny{ [.04,.11]} & \textbf{.84}\tiny{ [.79,.88]} & \textbf{.72}\tiny{ [.64,.79]} \\
\bottomrule
\end{tabular}
\caption{Main results on 150 human-annotated test cases with 95\% bootstrap CIs. AgentEval vs.\ Flat Step: $p < 0.001$ (FDRec, HA, RCA).}
\label{tab:main_results}
\end{table}

\textsc{AgentEval} achieves FDRec 0.89, representing $2.17\times$ higher recall than E2E (0.41) and outperforming Flat Step (0.67, $+$22 pp, $p < 0.001$). FDRec is computed against the 195 steps humans labeled as failing; for each method, the question is which of those 195 failing steps the method flags. By construction, E2E only inspects the final synthesis step, so failures originating earlier in the workflow are detectable only when they cause the final output to fail visibly. The $2.17\times$ gap therefore quantifies the recall lost to that structural restriction, not a difference in judge quality on equivalent inputs. Since the only difference from Flat Step is DAG structure (identical judges and rubrics), the $+22$ pp improvement over Flat Step further isolates the contribution of dependency modeling on top of step-level evaluation. Across all 450 test cases, 63\% of step-level failures are \emph{propagated} from upstream errors rather than locally caused, underscoring why dependency tracking is essential. Human agreement reaches $\kappa = 0.84$ [0.79, 0.88] (``almost perfect''), remaining strong on the failure subset ($\kappa = 0.76$, Appendix~\ref{sec:appendix_kappa_subsets}). Root cause accuracy of 72\% approaches the human ceiling ($\sim$81\%), with 72\% of incorrect attributions within 1 DAG hop of the true root cause (Appendix~\ref{sec:appendix_rca_errors}). Results are consistent across all three workflows (Appendix~\ref{sec:appendix_per_workflow}).

The DAG advantage holds on both typical ($+$23 pp FDRec) and edge cases ($+$19 pp), confirming the benefit is not driven by edge case oversampling (Appendix~\ref{sec:appendix_stratified}). The advantage grows with workflow length: $+$15 pp FDRec for $\leq$3 steps vs.\ $+$28 pp for $\geq$6 steps. On non-DAG traces ($\sim$12\%), performance degrades moderately (FDRec 0.82, RCA 0.58) but remains above flat evaluation.

\subsection{Ablation Study}
\label{sec:ablation}

\begin{table}[t]
\centering
\footnotesize
\begin{tabular}{@{}lccc@{}}
\toprule
\textbf{Configuration} & \textbf{FDRec}$\uparrow$ & \textbf{HA} ($\kappa$)$\uparrow$ & \textbf{RCA}$\uparrow$ \\
\midrule
Full \textsc{AgentEval} & \textbf{.89}\tiny{ [.84,.93]} & \textbf{.84}\tiny{ [.79,.88]} & \textbf{.72}\tiny{ [.64,.79]} \\
\quad $-$ DAG structure & .67\tiny{ [.61,.73]} & .71\tiny{ [.65,.76]} & .38\tiny{ [.30,.46]} \\
\quad \parbox[t]{2.9cm}{$-$ LLM judge\\[-1pt]{\tiny (replaced w/ rules)}} & .62\tiny{ [.56,.68]} & .66\tiny{ [.60,.72]} & .51\tiny{ [.43,.59]} \\
\quad $-$ Failure taxonomy & .82\tiny{ [.77,.87]} & .79\tiny{ [.74,.84]} & .54\tiny{ [.46,.62]} \\
\quad $-$ Calibration & .85\tiny{ [.80,.90]} & .76\tiny{ [.70,.81]} & .68\tiny{ [.60,.75]} \\
\bottomrule
\end{tabular}
\caption{Ablation study. DAG structure contributes the most ($-$22 pp FDRec, $-$34 pp RCA). All differences significant at $p < 0.01$.}
\label{tab:ablation}
\end{table}

Removing DAG structure causes the largest degradation ($-$22 pp FDRec, $-$34 pp RCA; $p < 0.001$), confirming dependency modeling as the most important design decision. Replacing LLM-as-judge with rules causes the second-largest drop ($-$27 pp FDRec). Removing the failure taxonomy primarily impacts RCA ($-$18 pp), and calibration anchors primarily affect human agreement ($-$8 pp $\kappa$).

Execution failures dominate across all workflows (42--55\%), but integration failures are most impactful: context loss has an average downstream amplification factor of 3.2$\times$, explaining why DAG-based evaluation outperforms flat evaluation. Results are robust across four judge models (GPT-4o, Claude~3.5~Sonnet, GPT-4o-mini, Llama~3~70B), with the DAG advantage remaining significant for all ($p < 0.01$; Appendix~\ref{sec:appendix_judge_sensitivity}). Same-family judging (Claude judging Claude agents) produces comparable or slightly \emph{lower} scores, ruling out self-evaluation inflation (Appendix~\ref{sec:appendix_judge_independence}).

\paragraph{Judge capability requirements.} The judge does not need to solve the agent's task in order to evaluate it: verifying that a tool selection is reasonable, that parameters match a schema, or that a synthesis step is grounded in retrieved context is in general easier than producing the action de novo. This asymmetry is what keeps the cost profile tractable, since a cheaper-than-agent judge can still rate steps usefully. Empirically, weaker judges preserve the DAG advantage in failure detection but degrade more sharply on root cause accuracy: GPT-4o-mini retains FDRec 0.83 against 0.89 for GPT-4o, but RCA drops from 0.72 to 0.65 (Appendix~\ref{sec:appendix_judge_sensitivity}). The practical implication is that the judge should be at least comparable to the weakest agent in the workflow on relevant verification tasks; deployments using judges substantially weaker than the agent should expect detection to remain useful while attribution becomes less reliable.

\paragraph{Counterfactual Validation.} To validate root cause identification, we replace identified root causes with gold-reference outputs and re-evaluate downstream steps. Across 30 sampled failing traces, counterfactual correction confirms \textsc{AgentEval}-identified root causes in 87\% of cases (26/30), with downstream scores improving by an average of 2.3 points. This procedure measures the combined effect of correcting the upstream output and the judge's sensitivity to the improved context that downstream steps then receive; it does not isolate either component. The engineering implication is that fixing the attributed root cause typically yields measurable improvement on downstream steps the agent itself would re-execute, which is precisely the intervention pattern engineers use during debugging. Cases where counterfactual correction does not improve downstream scores ($\sim$13\%) are flagged for manual review rather than auto-attributed.

\subsection{Regression Detection}

We simulate six model update scenarios across 18 scenario--workflow combinations. \textsc{AgentEval} achieves 88\% precision and 94\% recall for regression detection, correctly identifying no regression for benign prompt rephrasing and correctly localizing degradation for single-step tool deprecation. E2E evaluation requires 4$\times$ more test cases for equivalent statistical power. During the pilot, 23 regressions were detected organically (8 genuine, 12 borderline, 3 false positives). One illustrative case: a prompt template update for CS-Agent replaced a structured instruction with a free-form variant; \textsc{AgentEval} flagged a significant drop in \textsc{ParamGen} quality for the account-lookup step ($p < 0.01$, FDRec dropped 12~pp), which propagated to downstream steps, and the team reverted within hours. Additional case studies appear in Appendix~\ref{sec:appendix_regression_cases}.

\subsection{Cross-System Generalization}
\label{sec:generalization}

\begin{table}[t]
\centering
\footnotesize
\begin{tabular}{@{}llcc@{}}
\toprule
\textbf{Setting} & \textbf{Domain} & \textbf{FDRec} & \textbf{RCA} \\
\midrule
\multirow{3}{*}{Internal} & CS-Agent & .92 & .75 \\
& DA-Agent & .88 & .78 \\
& DP-Agent & .86 & .64 \\
\midrule
\multirow{2}{*}{$\tau$-bench} & Retail & .83\tiny{ [.76,.89]} & .61\tiny{ [.52,.70]} \\
& Airline & .79\tiny{ [.71,.86]} & .55\tiny{ [.45,.64]} \\
\midrule
SWE-bench & Code editing & .78\tiny{ [.69,.86]} & .52\tiny{ [.42,.62]} \\
\bottomrule
\end{tabular}
\caption{Cross-system generalization. Performance degrades on external benchmarks but remains above E2E baselines (FDRec 0.38 on $\tau$-bench, 0.35 on SWE-bench).}
\label{tab:cross_system}
\end{table}

To assess generalization, we evaluate on $\tau$-bench \citep{yao2024taubench} (120 traces) and SWE-bench \citep{jimenez2024swebench} (80 traces) without modifying the taxonomy or rubrics. \textsc{AgentEval} successfully parses 90\% of $\tau$-bench and 89\% of SWE-bench traces into valid DAGs. FDRec remains robust ($\geq$0.78), while RCA degrades more sharply ($-$14~pp $\tau$-bench, $-$20~pp SWE-bench), concentrated in novel failure patterns not in our taxonomy. To assess potential annotator bias, 2 independent annotators (graduate students, not authors) re-annotated 40 $\tau$-bench traces with comparable agreement ($\kappa = 0.74$ step quality), suggesting limited confirmation bias. This pattern of robust detection with degraded attribution on novel domains suggests the DAG-based approach generalizes as evaluation structure, while the taxonomy benefits from domain-specific extension.

\subsection{Architecture Scope Analysis}
\label{sec:scope_analysis}

Our internal agents exhibit a 12\% non-DAG rate. Simulation of increasing non-DAG rates (Appendix~\ref{sec:appendix_scope_detail}) shows that the DAG advantage remains significant up to $\sim$60\%, beyond which it becomes marginal ($<$5~pp). The boundary is not a property of the architecture alone but of how much execution structure can be reconstructed post hoc: retry loops are recoverable through unrolling, dynamic branches through timestamp resolution, and short reasoning loops through bounded iteration limits. Architectures that defeat reconstruction tend to share two properties: long unbounded reasoning loops, and parallel agent handoffs without explicit dependency declaration. \textsc{AgentEval} is therefore most useful for the sequential and moderately-branching tool-calling patterns that currently dominate production deployments; we view extension to cycle-aware and multi-agent settings as the natural next step (Section~\ref{sec:conclusion}).

\section{Deployment Experience}
\label{sec:deployment}

\textsc{AgentEval} operates as a sidecar service with OpenTelemetry-compatible trace collection ($<$2\% latency overhead). Evaluation runs asynchronously with a tiered judge fallback: GPT-4o primary, GPT-4o-mini secondary, and locally deployed Llama~3~70B as an API-independent fallback. Step-level evaluation takes $\sim$2 seconds per step; cost per trace is $\sim$\$0.02 with GPT-4o-mini. Multi-judge aggregation (3$\times$ cost) is recommended for release-gating only. At scale (100K traces/day), daily cost is $\approx$\$2K with GPT-4o-mini, motivating progressive evaluation where fast smoke tests gate full suites.

\begin{table}[t]
	\centering
	\resizebox{\columnwidth}{!}{%
		\footnotesize
		\setlength{\tabcolsep}{2pt}
		\begin{tabular}{@{}lrl@{}}
			\toprule
			\textbf{Metric} & \textbf{Value} & \textbf{Source} \\
			\midrule
			Total traces evaluated & 12,847 & System logs \\
			Unique evaluation runs & 342 & System logs \\
			Pre-release regressions detected & 23 & Alerts + triage \\
			\quad Genuine / Borderline / False & 8 / 12 / 3 & Team triage \\
			Config.\ changes from findings & 14 & Git commits \\
			Failure rate reduction (CS) & 31\%$\rightarrow$18\% & Eval suite \\
			Failure rate reduction (DA) & 27\%$\rightarrow$15\% & Eval suite \\
			Root-cause time (before) & 4.2 hr (med.) & Survey ($n$=47) \\
			Root-cause time (with AE) & 22 min (med.) & Logs ($n$=156) \\
			\bottomrule
		\end{tabular}%
	}
	\caption{Pilot deployment outcomes (4 months, 18 engineers). Note: baseline and pilot measurements use different instruments; the magnitude of improvement should be interpreted with caution (details in Appendix~\ref{sec:appendix_deployment_methodology}).}
	\label{tab:pilot_outcomes}
\end{table}

Over a 4-month pilot (September 2025--January 2026), three engineering teams used \textsc{AgentEval} to evaluate agents during active development (Table~\ref{tab:pilot_outcomes}). The most impactful outcome was the replacement of a previously manual root-cause investigation process with an automated propagation-aware analysis: median time to identify a root cause moved from 4.2 hours under the prior workflow (self-reported survey) to 22 minutes with \textsc{AgentEval} (system-logged). The two figures use different measurement instruments and the precise magnitude should be read with that caveat (Appendix~\ref{sec:appendix_deployment_methodology}); the directional improvement is consistent across all three teams. AgentEval's error propagation analysis revealed that 23\% of CS-Agent failures originated from a context truncation bug in the policy retrieval step; this single fix reduced CS-Agent's failure rate by 8 percentage points. For DA-Agent, taxonomy-guided analysis identified ``parameter hallucination'' (syntactically valid but semantically incorrect SQL WHERE clauses) as 31\% of execution failures; targeted few-shot examples reduced the parameter error rate from 27\% to 11\%. Practitioner feedback highlighted error propagation visibility as the primary benefit: ``The DAG visualization made error propagation chains immediately visible. We found three cascading failure patterns that end-to-end testing had missed entirely.'' (Senior Engineer). Key deployment lessons include the importance of taxonomy-first design and principled calibration anchor selection (Appendix~\ref{sec:appendix_lessons}). Onboarding a new workflow requires $\sim$20--30 person-hours, with same-domain workflows requiring less ($\sim$12--18 hrs) through partial reuse.

\begin{table}[t]
\centering
\footnotesize
\setlength{\tabcolsep}{1.8pt}
\begin{tabular}{@{}lccccc@{}}
\toprule
\textbf{Feature} & \rotatebox{60}{\textbf{\textsc{AE}}} & \rotatebox{60}{\textbf{LangSmith}} & \rotatebox{60}{\textbf{Arize}} & \rotatebox{60}{\textbf{Braintr.}} & \rotatebox{60}{\textbf{Inspect}} \\
\midrule
DAG dependency model & \cmark & \pmark & \xmark & \xmark & \xmark \\
Error propagation & \cmark & \xmark & \xmark & \xmark & \xmark \\
Root cause analysis & \cmark & \xmark & \xmark & \xmark & \xmark \\
Calibrated LLM judge & \cmark & \pmark & \pmark & \pmark & \cmark \\
Failure taxonomy & \cmark & \xmark & \xmark & \xmark & \xmark \\
CI/CD regression & \cmark & \cmark & \pmark & \cmark & \pmark \\
Open-source & Avail. & \cmark & \cmark & \cmark & \cmark \\
\bottomrule
\end{tabular}
\caption{Comparison with evaluation tools (public documentation, January 2026; features evolving rapidly). \cmark: full; \pmark: partial; \xmark: not supported.}
\label{tab:tool_comparison}
\end{table}

\section{Conclusion}
\label{sec:conclusion}

We presented \textsc{AgentEval}, an evaluation infrastructure that formalizes agent workflows as evaluation DAGs with step-level quality metrics, a hierarchical failure taxonomy, and automated regression testing. On three production workflows with predominantly sequential architectures, \textsc{AgentEval} achieves $2.17\times$ higher failure detection recall than end-to-end evaluation, $\kappa = 0.84$ human agreement, and 72\% root cause accuracy approaching the human ceiling (81\%). The ablation study confirms DAG-based dependency modeling as the single most impactful component, an advantage that grows with workflow complexity and holds across four judge models. Cross-system evaluation on $\tau$-bench and SWE-bench demonstrates transferability (FDRec $\geq 0.78$), while architecture scope analysis characterizes the practical boundary at $\sim$60\% non-DAG traces. A 4-month pilot validated practical utility, with specific AgentEval-informed fixes driving measurable failure rate reductions. Future work includes extending to multi-agent systems via cycle-aware evaluation graphs, automated taxonomy evolution through trace clustering, and integration of evaluation signals into agent training loops.

\paragraph{Reproducibility.} All materials are available at: \url{https://github.com/bettyguo/AgentEval}.

\section*{Acknowledgments}

We thank The University of Hong Kong and Stellaris AI Limited for their support of this work, and the anonymous reviewers for constructive feedback that improved the paper.

\section*{Limitations}

Our evaluation covers predominantly sequential agent architectures; the DAG advantage diminishes beyond $\sim$60\% non-DAG trace rates, limiting applicability to highly dynamic multi-agent systems (\S\ref{sec:scope_analysis}). All core results are from a single organization, though cross-system evaluation on two external benchmarks provides evidence of generalizability. LLM-as-judge evaluation introduces model-dependent biases; our cross-family design mitigates but does not eliminate these. The RCA algorithm is a practical heuristic, not formal causal inference. The deployment comparison uses different measurement instruments and should be interpreted with caution. We evaluate English-language workflows only. The commercial tool comparison is time-sensitive. Extended discussion of all limitations appears in Appendix~\ref{sec:appendix_limitations_extended}.

\section*{Ethical Considerations}

Our evaluation uses anonymized historical queries with no personally identifiable information; data handling was reviewed under our organization's internal data governance process. Human annotators were domain experts compensated at standard professional rates and informed of the study's purpose. LLM-as-judge evaluation introduces potential biases from the judge model (GPT-4o); we mitigate this through cross-family evaluation design but acknowledge residual model-dependent biases. The failure taxonomy was developed on internal data and may not capture failure modes relevant to all deployment contexts.

\bibliography{references}

\appendix

\section{Non-DAG Trace Handling}
\label{sec:appendix_nondag}

Approximately 12\% of traces involve patterns that do not conform to strict DAGs, primarily retry loops (8\%) and dynamic branching (4\%). Retry loops are handled by \emph{unrolling} into sequential attempts within the DAG, preserving the most recent attempt's output as the step's canonical result. Dynamic branches are resolved using trace timestamps to reconstruct the executed path. For traces where DAG reconstruction fails entirely (0.8\%), the system falls back to flat step evaluation and flags the trace for manual review.

\section{Full Failure Taxonomy}
\label{sec:appendix_taxonomy}

Table~\ref{tab:full_taxonomy} presents the complete three-level failure taxonomy with Level~3 subcategories.

\paragraph{Construction methodology.} The taxonomy was developed through affinity diagramming. Three authors independently coded the same 523 development traces (entirely disjoint from the 450 evaluation test cases), generating an initial pool of failure descriptors. These descriptors were then merged into candidate categories through multiple consensus sessions. The resulting 9 Level~2 categories and 21 Level~3 subcategories were locked before any evaluation data was annotated; the evaluation set was used only to confirm that the categories were comprehensive enough to label observed failures, never to refine the taxonomy itself. During the pilot deployment, three additional subcategories were proposed and merged through a propose-discuss-merge governance protocol (Appendix~\ref{sec:appendix_lessons}); these additions are reflected in the version reported here.

\begin{table*}[ht]
\centering
\footnotesize
\begin{tabular}{@{}lllp{5.5cm}@{}}
\toprule
\textbf{Level 1} & \textbf{Level 2} & \textbf{Level 3} & \textbf{Description \& Example} \\
\midrule
\multirow{7}{*}{Planning} & \multirow{2}{*}{Goal misinterpretation} & Scope error & Agent addresses wrong aspect of the query \\
 & & Ambiguity failure & Agent selects wrong interpretation without clarification \\
\cmidrule{2-4}
 & \multirow{3}{*}{Missing steps} & Tool omission & Required tool not included in plan \\
 & & Verification gap & Plan lacks result verification \\
 & & Prerequisite skip & Dependency step omitted \\
\cmidrule{2-4}
 & \multirow{2}{*}{Incorrect ordering} & Dependency violation & Step executed before prerequisites \\
 & & Suboptimal sequence & Valid but inefficient ordering \\
\midrule
\multirow{7}{*}{Execution} & \multirow{2}{*}{Wrong tool selection} & Category error & Wrong tool category \\
 & & Granularity mismatch & Correct category but wrong specificity \\
\cmidrule{2-4}
 & \multirow{3}{*}{Parameter errors} & Type mismatch & Wrong data type \\
 & & Value error & Correct type but incorrect value \\
 & & Missing required & Required parameter omitted \\
\cmidrule{2-4}
 & \multirow{2}{*}{API/tool failures} & Timeout & Tool call exceeds time limit \\
 & & Error response & Tool returns error, agent fails to handle \\
\midrule
\multirow{6}{*}{Integration} & \multirow{2}{*}{Context loss} & Truncation & Information dropped between steps \\
 & & Selective omission & Agent ignores relevant output parts \\
\cmidrule{2-4}
 & \multirow{2}{*}{Output hallucination} & Fabrication & Information not grounded in any result \\
 & & Conflation & Information from different steps incorrectly merged \\
\cmidrule{2-4}
 & \multirow{2}{*}{Premature termination} & Partial completion & Agent declares success prematurely \\
 & & Loop exit & Agent exits retry loop too early \\
\bottomrule
\end{tabular}
\caption{Complete three-level failure taxonomy with 21 Level~3 subcategories.}
\label{tab:full_taxonomy}
\end{table*}

\section{LLM-as-Judge Prompt Templates}
\label{sec:appendix_prompts}

All prompts use GPT-4o (\texttt{gpt-4o-2024-08-06}) with temperature $= 0$ and \texttt{max\_tokens=1024}.

\subsection{Tool Selection (\textsc{ToolSel})}

\smallskip
\noindent\fbox{\parbox{0.95\columnwidth}{\footnotesize
\textbf{System}: You are an expert evaluator assessing the quality of an AI agent's tool selection in a multi-step workflow. Evaluate strictly according to the rubric below. First provide step-by-step reasoning, then your score on a separate line as ``Score: N''.

\textbf{Task}: Given the agent's current context and the tool it selected, rate the tool selection quality on a 1--5 scale.

\textbf{Rubric}:
\begin{itemize}[nosep,leftmargin=1em]
\item[5:] \emph{Optimal}. Best available tool for the subtask.
\item[4:] \emph{Acceptable}. Valid tool but better alternative exists.
\item[3:] \emph{Partially correct}. Correct category but suboptimal tool.
\item[2:] \emph{Poor}. Wrong category but may produce some useful output.
\item[1:] \emph{Incorrect}. Cannot accomplish the subtask.
\end{itemize}

\textbf{Calibration Examples}: [5 anchor examples omitted for space]

\textbf{Context}: \{agent\_context\}\\
\textbf{Available tools}: \{tool\_list\}\\
\textbf{Selected tool}: \{agent\_tool\_selection\}\\
\textbf{Subtask description}: \{subtask\}
}}

\subsection{Planning (\textsc{Plan})}

\smallskip
\noindent\fbox{\parbox{0.95\columnwidth}{\footnotesize
\textbf{System}: You are an expert evaluator assessing the quality of an AI agent's execution plan. First provide reasoning, then ``Score: N''.

\textbf{Task}: Rate the plan quality on a 1--5 scale given the user query and available tools.

\textbf{Rubric}:
\begin{itemize}[nosep,leftmargin=1em]
\item[5:] \emph{Complete \& optimal}. All subtasks identified, feasible, correctly ordered.
\item[4:] \emph{Mostly complete}. Minor omission but critical steps present.
\item[3:] \emph{Partial}. Covers main task but misses 1--2 important subtasks.
\item[2:] \emph{Inadequate}. Major gaps or infeasible steps.
\item[1:] \emph{Incorrect}. Plan would not address the user query.
\end{itemize}

\textbf{Calibration Examples}: [omitted for space]

\textbf{User query}: \{query\}\\
\textbf{Available tools}: \{tool\_list\}\\
\textbf{Agent plan}: \{agent\_plan\}
}}

\section{Human Agreement by Score Subset}
\label{sec:appendix_kappa_subsets}

\begin{table}[H]
\centering
\footnotesize
\begin{tabular}{@{}lrc@{}}
\toprule
\textbf{Subset} & \textbf{$n$ (steps)} & \textbf{$\kappa$} \\
\midrule
All steps & 987 & .84 \\
Failing (score $\leq$ 2) & 195 & .76 \\
Borderline (score $=$ 3) & 146 & .69 \\
Passing (score $\geq$ 4) & 646 & .87 \\
\bottomrule
\end{tabular}
\caption{Human agreement by score subset. Agreement remains substantial on failures ($\kappa \geq 0.69$), confirming the overall $\kappa = 0.84$ is not inflated by majority-class agreement.}
\label{tab:kappa_subsets_table}
\end{table}

\section{RCA Error Analysis}
\label{sec:appendix_rca_errors}

Among the 28\% of incorrect attributions, three patterns dominate: (1) \emph{multi-branch convergence} (41\% of RCA errors), where independent upstream failures converge at a single step; (2) \emph{long propagation chains} (33\%), where the true root cause is $\geq$3 steps upstream; and (3) \emph{threshold boundary cases} (26\%), where the root cause scores marginally above the threshold.

The median distance between attributed and true root cause for incorrect attributions is 1 hop, with 72\% of errors within 1 hop and 91\% within 2 hops. This means most errors send engineers to a \emph{nearby} step rather than a distant one, substantially reducing misattribution cost compared to random assignment (mean distance 2.8 hops).

\section{RCA Attribution Strategy Comparison}
\label{sec:appendix_rca_strategies}

\begin{table}[H]
\centering
\footnotesize
\begin{tabular}{@{}lcc@{}}
\toprule
\textbf{Attribution Strategy} & \textbf{RCA Acc.} & \textbf{False Attr.} \\
\midrule
Greedy lowest-parent (default) & .72 & .06 \\
Full-path minimum & .75 & .14 \\
Weighted propagation & .73 & .09 \\
\bottomrule
\end{tabular}
\caption{Comparison of RCA strategies. Full-path minimum improves accuracy by 3~pp but doubles false attributions to distant ancestors.}
\label{tab:rca_strategies}
\end{table}

\paragraph{Tradeoff discussion.} The three strategies differ in how aggressively they search upstream for a root cause when multiple parents have low scores. Greedy lowest-parent stops at the immediate parent with the lowest quality score. Full-path minimum walks the entire ancestor chain and selects the globally lowest-scoring node. Weighted propagation interpolates between the two by weighting parent scores by edge confidence.

Full-path minimum achieves a 3~pp RCA improvement, but at the cost of more than doubling the false-attribution rate (.06 to .14) by reaching for distant ancestors that may have been transient or already compensated for by intermediate steps. During pilot use, engineers reported that distant misattributions were disruptive: the wasted investigation time on a misattribution three hops upstream consistently exceeded the time saved by the 3~pp accuracy gain on cases where a distant attribution was correct. We therefore selected greedy lowest-parent as the default. Weighted propagation offers a smaller intermediate tradeoff (1~pp RCA improvement, +3~pp false attribution) and may suit deployments with reliable edge-confidence signals, though our pilot did not require it.

\section{Stratified Results}
\label{sec:appendix_stratified}

\begin{table}[H]
\centering
\footnotesize
\setlength{\tabcolsep}{2.5pt}
\begin{tabular}{@{}llccc@{}}
\toprule
\textbf{Subset} & \textbf{Method} & \textbf{FDRec} & \textbf{HA ($\kappa$)} & \textbf{RCA} \\
\midrule
\multirow{2}{*}{Typical ($\sim$80\%)} & Flat Step & .64 & .70 & .36 \\
& \textsc{AgentEval} & \textbf{.87}\tiny{ [.81,.92]} & \textbf{.83} & \textbf{.70}\tiny{ [.61,.78]} \\
\midrule
\multirow{2}{*}{Edge ($\sim$20\%)} & Flat Step & .76 & .74 & .44 \\
& \textsc{AgentEval} & \textbf{.95}\tiny{ [.89,.98]} & \textbf{.87} & \textbf{.78}\tiny{ [.67,.87]} \\
\bottomrule
\end{tabular}
\caption{Results by case type. DAG advantage holds on both subsets.}
\label{tab:stratified_results}
\end{table}

\begin{table}[H]
\centering
\footnotesize
\begin{tabular}{@{}lccc@{}}
\toprule
\textbf{Steps} & \textbf{$\Delta$FDRec} & \textbf{$\Delta$RCA} & \textbf{$n$ traces} \\
\midrule
$\leq$3 & +15~pp & +22~pp & 127 \\
4--5 & +21~pp & +33~pp & 198 \\
$\geq$6 & +28~pp & +42~pp & 125 \\
\bottomrule
\end{tabular}
\caption{DAG advantage by workflow length.}
\label{tab:length_stratified}
\end{table}

\section{Per-Workflow Breakdown}
\label{sec:appendix_per_workflow}

\begin{table}[H]
\centering
\footnotesize
\setlength{\tabcolsep}{2.8pt}
\begin{tabular}{@{}llcccc@{}}
\toprule
\textbf{Wkfl} & \textbf{Method} & \textbf{FDRec} & \textbf{FPR} & \textbf{HA} & \textbf{RCA} \\
\midrule
\multirow{4}{*}{\rotatebox{90}{\scriptsize CS}} & E2E & .35 & .07 & .48 & N/A \\
& Flat & .63 & .17 & .68 & .33 \\
& Rule & .54 & .04 & .60 & .41 \\
& \textsc{AE} & \textbf{.92} & .08 & \textbf{.86} & \textbf{.75} \\
\midrule
\multirow{4}{*}{\rotatebox{90}{\scriptsize DA}} & E2E & .43 & .09 & .54 & N/A \\
& Flat & .69 & .14 & .73 & .42 \\
& Rule & .60 & .05 & .65 & .49 \\
& \textsc{AE} & \textbf{.88} & .07 & \textbf{.85} & \textbf{.78} \\
\midrule
\multirow{4}{*}{\rotatebox{90}{\scriptsize DP}} & E2E & .46 & .08 & .55 & N/A \\
& Flat & .70 & .13 & .72 & .40 \\
& Rule & .61 & .06 & .64 & .46 \\
& \textsc{AE} & \textbf{.86} & .06 & \textbf{.81} & \textbf{.64} \\
\bottomrule
\end{tabular}
\caption{Per-workflow breakdown.}
\label{tab:per_workflow}
\end{table}

Performance on non-DAG traces ($\sim$12\%, 54 traces total): FDRec 0.82 [0.73, 0.90] and RCA 0.58 [0.46, 0.69], lower than overall results but above Flat Step (FDRec 0.64, RCA 0.35 on the same subset). Degradation concentrates in retry-loop traces; dynamic branching traces show smaller degradation (FDRec 0.85).

\section{Judge Model Sensitivity}
\label{sec:appendix_judge_sensitivity}

\begin{table}[H]
\centering
\footnotesize
\begin{tabular}{@{}lcccc@{}}
\toprule
\textbf{Judge Model} & \textbf{FDRec} & \textbf{FPR} & \textbf{HA} ($\kappa$) & \textbf{RCA} \\
\midrule
GPT-4o & \textbf{.89} & .07 & \textbf{.84} & \textbf{.72} \\
Claude 3.5 Sonnet & .87 & .09 & .81 & .69 \\
GPT-4o-mini & .83 & .11 & .77 & .65 \\
Llama 3 70B & .80 & .13 & .73 & .61 \\
\bottomrule
\end{tabular}
\caption{Judge sensitivity. DAG advantage over Flat Step remains significant ($p < 0.01$) for all judges.}
\label{tab:judge_sensitivity}
\end{table}

\section{Agent--Judge Independence}
\label{sec:appendix_judge_independence}

\begin{table}[H]
\centering
\footnotesize
\setlength{\tabcolsep}{2.5pt}
\begin{tabular}{@{}llccc@{}}
\toprule
\textbf{Workflow} & \textbf{Judge} & \textbf{FDRec} & \textbf{HA} & \textbf{RCA} \\
\midrule
\multirow{2}{*}{CS (Claude agent)} & GPT-4o & .92 & .86 & .75 \\
 & Claude 3.5 Sonnet & .90 & .83 & .73 \\
\midrule
\multirow{2}{*}{DA (Claude agent)} & GPT-4o & .88 & .85 & .78 \\
 & Claude 3.5 Sonnet & .86 & .82 & .75 \\
\midrule
\multirow{2}{*}{DP (Llama agent)} & GPT-4o & .86 & .81 & .64 \\
 & Claude 3.5 Sonnet & .85 & .80 & .63 \\
\bottomrule
\end{tabular}
\caption{Same-family judging produces comparable or lower scores, ruling out self-evaluation inflation.}
\label{tab:judge_independence}
\end{table}

\section{Architecture Scope Details}
\label{sec:appendix_scope_detail}

To characterize the practical boundary of DAG-based evaluation, we simulate increasing non-DAG rates by randomly converting DAG traces to non-DAG (loop/branch) structures. At 30\% non-DAG rate, FDRec remains 0.83 and RCA 0.63; at 50\%, FDRec is 0.78 and RCA 0.55, still above flat evaluation but with diminished advantage. The practical boundary is approximately 60\% non-DAG rate, beyond which the DAG advantage becomes marginal ($<$5~pp). \textsc{AgentEval} is most valuable for sequential and moderately-branching architectures, which represent the majority of current production deployments (Table~\ref{tab:nondag_rates}).

\begin{table}[H]
	\centering
	\resizebox{\columnwidth}{!}{%
		\footnotesize
		\begin{tabular}{@{}lcc@{}}
			\toprule
			\textbf{Architecture Class} & \textbf{Est.\ Non-DAG \%} & \textbf{Source} \\
			\midrule
			Sequential tool-calling & 8--15\% & Internal agents \\
			Moderate branching & 20--30\% & $\tau$-bench \\
			SWE agents (edit-test loops) & 25--35\% & SWE-bench \\
			Multi-agent collaboration & 40--60\% & Estimated\textsuperscript{$\dagger$} \\
			Tree-of-thought / LATS & 60--80\% & Estimated\textsuperscript{$\dagger$} \\
			\bottomrule
		\end{tabular}%
	}
	\caption{Estimated non-DAG trace rates. \textsuperscript{$\dagger$}Based on architectural analysis, not empirically validated.}
	\label{tab:nondag_rates}
\end{table}

\section{Regression Case Studies}
\label{sec:appendix_regression_cases}

Three illustrative cases from the pilot: (1) A prompt template update for CS-Agent replaced a structured instruction with a free-form variant. \textsc{AgentEval} flagged a drop in \textsc{ParamGen} quality for account-lookup ($p < 0.01$, FDRec dropped 12~pp), which propagated downstream. The team reverted within hours. (2) DA-Agent's SQL generation degraded after an LLM provider-side update; \textsc{AgentEval} localized the regression to code generation steps, enabling a targeted fix. (3) A DP-Agent schema change improved average-case performance by 2~pp but degraded long-document handling by 11~pp, detected only by the edge-case tier of progressive evaluation.

\section{Deployment Methodology Details}
\label{sec:appendix_deployment_methodology}

\emph{Baseline}: Self-reported surveys from 12 engineers covering 47 incidents before adoption. Median: 4.2 hours; P25: 1.8 hours; P75: 8.5 hours; P90: 14 hours. Self-reports may overestimate (recall bias) and capture total time including communication overhead.

\emph{With \textsc{AgentEval}}: System-logged times from 156 events. Median: 22 minutes; P25: 8 minutes; P75: 48 minutes; P90: 1.6 hours. Logged times capture alert-to-first-code-change interval and may \emph{underestimate} by excluding initial investigation.

Both measurements consistently show substantial improvement across all three teams, but the exact magnitude should be interpreted with caution given measurement asymmetry and confounds (failure severity, Hawthorne effects, concurrent improvements).

Additional deployment impact: For DA-Agent, taxonomy-guided analysis identified ``parameter hallucination'' (syntactically valid but semantically incorrect SQL WHERE clauses) as 31\% of execution failures. Targeted few-shot examples reduced DA-Agent's parameter error rate from 27\% to 11\%.

Onboarding a new workflow requires $\sim$20--30 person-hours. Same-domain workflows require less ($\sim$12--18 hrs) through partial reuse. Preliminary experiments suggest GPT-4o-assisted DAG schema proposals reduce setup time by approximately 40\%.

\section{Deployment Lessons Learned}
\label{sec:appendix_lessons}

Five insights from deployment:

\emph{First, taxonomy-first design is essential.} After the first month, 34\% of CS-Agent failures were ``parameter hallucination'' errors not captured by initial metrics. \emph{Lesson}: Start with failure modes, not metrics.

\emph{Second, calibration is critical and brittle.} Uncalibrated judges showed 15\% lower agreement. Five stratified anchors per metric was the minimum; random anchors underperformed due to skewed score distributions.

\emph{Third, schema-defined and trace-inferred DAGs serve complementary purposes.} Structural deviation proved a useful quality signal: deviant traces were associated with 2.1$\times$ higher failure rates.

\emph{Fourth, progressive evaluation changes behavior.} Teams that adopted smoke tests ran 3.4$\times$ more evaluation cycles per sprint.

\emph{Fifth, taxonomy maintenance is a sociotechnical challenge.} We added 3 subcategories during the pilot and adopted a ``propose-discuss-merge'' protocol for governance.

\section{Extended Limitations}
\label{sec:appendix_limitations_extended}

\paragraph{Methodological Boundaries.} Our evaluation dataset is sourced from anonymized logs using production tools and prompts but does not reflect full live-system complexity. The reported $\kappa = 0.84$ reflects agreement against majority-voted references; per-annotator agreement ($\kappa = 0.78$--$0.86$, Table~\ref{tab:per_annotator_agreement}) provides a more conservative estimate. The 150 human-annotated cases provide adequate power for aggregate metrics but wide per-workflow CIs. Ground truth for internal evaluation was produced by experts separate from authors; $\tau$-bench and SWE-bench annotations were produced by authors with independent validation on a subset. Cost analysis assumes January 2026 API pricing.

\paragraph{Generalization Scope.} Our taxonomy was derived from three workflow types and may not cover other domains; cross-system evaluation suggests 76\% taxonomy coverage on out-of-domain agents. We evaluate English-language workflows only. The 5-type step classification covers common tool-calling patterns but may need extension for specialized architectures.

\section{Threshold Sensitivity Analysis}
\label{sec:appendix_threshold}

\begin{table}[H]
\centering
\footnotesize
\begin{tabular}{@{}lccc@{}}
\toprule
\textbf{Threshold shift} & \textbf{FDRec} & \textbf{FPR} & \textbf{RCA} \\
\midrule
$\theta - 0.5$ (more sensitive) & .92 & .14 & .69 \\
$\theta$ (default) & .89 & .07 & .72 \\
$\theta + 0.5$ (more conservative) & .85 & .04 & .67 \\
\bottomrule
\end{tabular}
\caption{Sensitivity to uniform threshold perturbation ($\pm 0.5$).}
\label{tab:threshold_sensitivity}
\end{table}

Per-type thresholds: $\theta_{\textsc{Plan}} = 3.0$, $\theta_{\textsc{ToolSel}} = 3.0$, $\theta_{\textsc{ParamGen}} = 2.5$, $\theta_{\textsc{Exec}} = 3.0$, $\theta_{\textsc{Synth}} = 3.0$, selected by grid search on a held-out 52-trace subset. \textsc{ParamGen} thresholds are most sensitive: $\pm 0.5$ shifts FDRec by $\pm$4~pp, compared to $\pm$2~pp for other types.

\paragraph{Aggregation Function.} The workflow-level score is the weighted harmonic mean:
\begin{equation}
Q(\mathcal{G}) = \frac{\sum_{i} w_i}{\sum_{i} w_i / q(v_i)}, \quad w_i = |\text{desc}(v_i)| + 1
\end{equation}
where $\text{desc}(v_i)$ is the set of descendants of $v_i$.

\section{Counterfactual Analysis}
\label{sec:appendix_counterfactual}

To demonstrate counterfactual analysis, we replace a CS-Agent root cause ($v_2$, tool selection) with the gold-reference output and re-evaluate downstream. Scores improve: $v_3$ from 2.1 to 4.6, $v_4$ from 1.8 to 4.3, $v_5$ from 2.3 to 4.1. Across 30 sampled failing traces, counterfactual analysis confirms \textsc{AgentEval}-identified root causes in 87\% of cases. This measures the combined effect of error correction and judge sensitivity to improved context.

\section{Systematic Disagreement Patterns}
\label{sec:appendix_disagreement}

Analyzing 156 step evaluations where \textsc{AgentEval} disagrees with human majority vote: (1) \emph{partial correctness} (43\%): steps with partially correct output that humans rate leniently but the judge rates strictly, especially \textsc{Synth} steps; (2) \emph{context sensitivity} (31\%): cases where human evaluators use external knowledge not in the trace; (3) \emph{borderline thresholds} (26\%): steps near the pass/fail boundary where judgments diverge by exactly 1 point.

\section{Parsing Failure Analysis}
\label{sec:appendix_parsing}

Of 21 total parsing failures (12 $\tau$-bench + 9 SWE-bench): multi-agent handoffs without explicit dependency (8 traces), unbounded loops exceeding the 5-iteration unrolling limit (7 traces), and deeply nested conditional branches (6 traces).

\section{Error Propagation Statistics}
\label{sec:appendix_propagation}

Across all 450 test cases, 847 total step-level failures: 312 (36.8\%) root cause failures, 535 (63.2\%) propagated failures. Average propagation chain length: 2.1 steps. Context loss failures have the longest average chain (3.2 steps); API/tool failures the shortest (1.4 steps).

\section{Per-Annotator Agreement}
\label{sec:appendix_annotator}

\begin{table}[H]
\centering
\footnotesize
\begin{tabular}{@{}lcc@{}}
\toprule
\textbf{Annotator Pair} & \textbf{$\kappa$ (vs.\ AgentEval)} & \textbf{$\kappa$ (pairwise)} \\
\midrule
Annotator 1 (CS) & 0.86 & 0.81 \\
Annotator 2 (CS) & 0.83 & 0.78 \\
Annotator 3 (CS) & 0.82 & 0.77 \\
Annotator 4 (DA) & 0.85 & 0.80 \\
Annotator 5 (DA) & 0.84 & 0.79 \\
Annotator 6 (DA) & 0.81 & 0.76 \\
Annotator 7 (DP) & 0.80 & 0.78 \\
Annotator 8 (DP) & 0.78 & 0.75 \\
Annotator 9 (DP) & 0.79 & 0.77 \\
\bottomrule
\end{tabular}
\caption{Per-annotator agreement.}
\label{tab:per_annotator_agreement}
\end{table}

\section{Per-Metric Human Agreement}
\label{sec:appendix_per_metric}

\begin{table}[H]
\centering
\footnotesize
\begin{tabular}{@{}llc@{}}
\toprule
\textbf{Step Type} & \textbf{Metric} & \textbf{$\kappa$} \\
\midrule
\textsc{Plan} & Completeness & 0.78 \\
\textsc{Plan} & Feasibility & 0.81 \\
\textsc{ToolSel} & Selection accuracy & 0.92 \\
\textsc{ToolSel} & Tool relevance & 0.85 \\
\textsc{ParamGen} & Param.\ correctness & 0.89 \\
\textsc{ParamGen} & Param.\ completeness & 0.87 \\
\textsc{Synth} & Faithfulness & 0.80 \\
\textsc{Synth} & Completeness & 0.76 \\
\textsc{Synth} & Coherence & 0.82 \\
\midrule
\emph{Average} & & \emph{0.83} \\
\bottomrule
\end{tabular}
\caption{Per-metric human agreement (single-judge GPT-4o).}
\label{tab:per_metric_agreement}
\end{table}

\section{Score Distribution and Confusion Matrix}
\label{sec:appendix_confusion}

\begin{table}[H]
\centering
\footnotesize
\begin{tabular}{@{}lccccc@{}}
\toprule
\textbf{Score} & \textbf{1} & \textbf{2} & \textbf{3} & \textbf{4} & \textbf{5} \\
\midrule
Human (\%) & 8.2 & 11.5 & 14.8 & 28.7 & 36.8 \\
AgentEval (\%) & 7.6 & 11.5 & 16.3 & 29.4 & 35.2 \\
\bottomrule
\end{tabular}
\caption{Score distribution across 150 human-annotated cases (step-level).}
\label{tab:score_distribution}
\end{table}

\begin{table}[H]
\centering
\footnotesize
\setlength{\tabcolsep}{3.5pt}
\begin{tabular}{@{}l|ccccc|r@{}}
\toprule
\multirow{2}{*}{\textbf{Human}} & \multicolumn{5}{c|}{\textbf{AgentEval}} & \multirow{2}{*}{\textbf{Total}} \\
& \textbf{1} & \textbf{2} & \textbf{3} & \textbf{4} & \textbf{5} & \\
\midrule
\textbf{1} & 71 & 8 & 2 & 0 & 0 & 81 \\
\textbf{2} & 4 & 93 & 15 & 2 & 0 & 114 \\
\textbf{3} & 0 & 11 & 112 & 23 & 0 & 146 \\
\textbf{4} & 0 & 2 & 28 & 228 & 25 & 283 \\
\textbf{5} & 0 & 0 & 4 & 37 & 322 & 363 \\
\midrule
\textbf{Total} & 75 & 114 & 161 & 290 & 347 & 987 \\
\bottomrule
\end{tabular}
\caption{Confusion matrix (987 step evaluations). Binary pass/fail agreement (threshold 3.0) is 94.0\%.}
\label{tab:confusion_matrix}
\end{table}

\section{DAG Example Figure}
\label{sec:appendix_dag_example}

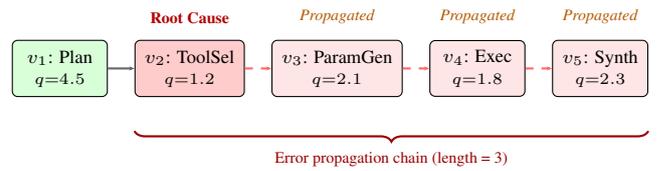
\begin{figure}[H]
\centering
\begin{tikzpicture}[
  node distance=0.55cm and 0.25cm,
  step/.style={rectangle, draw, rounded corners=2pt, minimum height=0.55cm, minimum width=1.25cm, font=\scriptsize, align=center, fill=#1},
  scorelabel/.style={font=\tiny, text=#1},
  depline/.style={-{Stealth[length=2pt]}, thick, black!60},
  propline/.style={-{Stealth[length=2pt]}, thick, red!60, dashed},
]
\node[step=green!15] (v1) {$v_1$: Plan\\{\tiny $q{=}4.5$}};
\node[step=red!20, right=0.35cm of v1] (v2) {$v_2$: ToolSel\\{\tiny $q{=}1.2$}};
\node[step=red!10, right=0.35cm of v2] (v3) {$v_3$: ParamGen\\{\tiny $q{=}2.1$}};
\node[step=red!10, right=0.35cm of v3] (v4) {$v_4$: Exec\\{\tiny $q{=}1.8$}};
\node[step=red!10, right=0.35cm of v4] (v5) {$v_5$: Synth\\{\tiny $q{=}2.3$}};
\node[font=\tiny\bfseries, text=red!70!black, above=0.08cm of v2] {Root Cause};
\node[font=\tiny\itshape, text=orange!70!black, above=0.08cm of v3] {Propagated};
\node[font=\tiny\itshape, text=orange!70!black, above=0.08cm of v4] {Propagated};
\node[font=\tiny\itshape, text=orange!70!black, above=0.08cm of v5] {Propagated};
\draw[depline] (v1) -- (v2);
\draw[propline] (v2) -- (v3);
\draw[propline] (v3) -- (v4);
\draw[propline] (v4) -- (v5);
\draw[decorate, decoration={brace, amplitude=4pt, mirror}, thick, red!60!black]
  ([yshift=-0.45cm]v2.south west) -- ([yshift=-0.45cm]v5.south east)
  node[midway, below=4pt, font=\tiny, text=red!60!black] {Error propagation chain (length = 3)};
\end{tikzpicture}
\caption{DAG-based evaluation of a CS-Agent workflow. A wrong tool selection at $v_2$ (root cause, $q=1.2$) propagates through $v_3$--$v_5$.}
\label{fig:dag_example}
\end{figure}

\section{Failure Mode Distribution}
\label{sec:appendix_failure_modes}

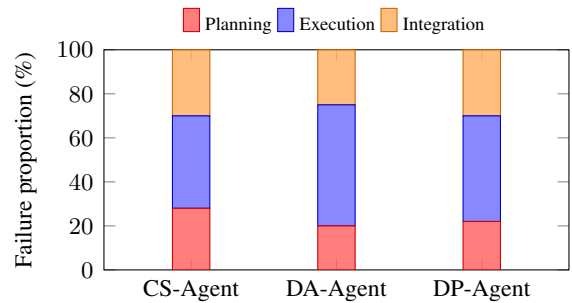
\begin{figure}[H]
\centering
\begin{tikzpicture}
\begin{axis}[
  ybar stacked,
  width=\columnwidth,
  height=4.5cm,
  bar width=14pt,
  ymin=0, ymax=100,
  ylabel={\footnotesize Failure proportion (\%)},
  symbolic x coords={CS-Agent, DA-Agent, DP-Agent},
  xtick=data,
  x tick label style={font=\footnotesize},
  y tick label style={font=\footnotesize},
  ylabel style={font=\footnotesize},
  legend style={font=\scriptsize, at={(0.5,1.02)}, anchor=south, legend columns=3, draw=none},
  enlarge x limits=0.3,
  every axis plot/.append style={fill opacity=0.85},
]
\addplot[fill=red!60!white, draw=red!80!black] coordinates {(CS-Agent,28) (DA-Agent,20) (DP-Agent,22)};
\addplot[fill=blue!55!white, draw=blue!80!black] coordinates {(CS-Agent,42) (DA-Agent,55) (DP-Agent,48)};
\addplot[fill=orange!60!white, draw=orange!80!black] coordinates {(CS-Agent,30) (DA-Agent,25) (DP-Agent,30)};
\legend{Planning, Execution, Integration}
\end{axis}
\end{tikzpicture}
\caption{Failure mode distribution across workflows.}
\label{fig:failure_analysis}
\end{figure}

\end{document}